\documentclass[a4paper,traditabstract]{aa} 
\usepackage{graphicx,txfonts,astro,natbib,booktabs}
\usepackage{epstopdf,enumitem,hyperref,xcolor}
\DeclareGraphicsExtensions{.pdf,.png,.eps,.gif,.jpg,.ps}
\begin{document}
\def\mvir{\ensuremath{M_{\rm vir}}}
\def\mtot{\ensuremath{M_{\rm tot}}}
\def\tdust{\ensuremath{T_{\rm dust}}}
\def\mc#1{\multicolumn{2}{c}{#1}}
\def\mmc#1#2{\multicolumn{#1}{c}{#2}}
\def\snupeak{\ensuremath{S_\nu^{\rm peak}}}
\def\snuint{\ensuremath{S_\nu^{\rm int}}}
\def\peakNh{\ensuremath{N_\h^{\rm peak}}}
\def\avNh{\ensuremath{\aver{N_\h}}}
\def\peaknh{\ensuremath{n_\h^{\rm peak}}}
\def\avnh{\ensuremath{\aver{n_\h}}}
\def\flux{\ensuremath{S_{870}}}
\def\pflux{\ensuremath{S_{870}^{\rm peak}}}
\def\iflux{\ensuremath{S_{870}^{\rm int}}}
\def\Rflux{\ensuremath{R_{\rm flux}}}
\def\Rgeom{\ensuremath{R_{\rm geom}}}
\def\phb#1{\par\textbf{**** #1 ****}}
\def\tfm#1{\tablefootmark{(#1)}}
\def\tft#1#2{{\\\tablefoottext{#1}{#2}}}
\def\tfoot#1{{\tablefoot{#1}}}

\title{The upGREAT 1.9 THz multi-pixel high resolution spectrometer for the SOFIA Observatory}

\author{C. Risacher\inst{1}, R. G\"usten\inst{1}, J. Stutzki\inst{2}, H.-W. H\"ubers\inst{3}
  A. Bell\inst{1}, C. Buchbender\inst{2},  D. B\"uchel\inst{2}, T. Csengeri\inst{1}, 
  U. U. Graf\inst{2}, S. Heyminck\inst{1}, R. D. Higgins\inst{2}, C. E. Honingh\inst{2}, K. Jacobs\inst{2}, 
  B. Klein\inst{1,5}, Y. Okada\inst{2}, A. Parikka\inst{2}, P. P\"utz\inst{2}, N. Reyes\inst{1,4}, O. Ricken\inst{1}, 
  D. Riquelme\inst{1}, R. Simon\inst{2} \and H. Wiesemeyer\inst{1}
}

\offprints{crisache@mpifr.de}

\institute{%
  Max-Planck-Institut f\"ur Radioastronomie, Auf dem H\"ugel 69, 53121, Bonn, Germany
  \and
  I. Physikalisches Institut der Universit\"at zu K\"oln, Z\"ulpicher Strasse 77, 50937 K\"oln, Germany
\and 
  Institute of Planetary Research, German Aerospace Center (DLR),
Rutherfordstr. 2, 12489 Berlin, Germany 
   \and
Departamento de Ingenier\'{i}a El\'ectrica, Universidad de Chile, Santiago, Chile
  \and
  University of Applied Sciences Bonn-Rhein-Sieg,
Sankt Augustin, 53757 Germany
}

\date{Received April 30, 2016}

\abstract{We present a new multi-pixel high resolution (R $\gtrsim$ $10^7$) spectrometer for the Stratospheric Observatory for Far-Infrared Astronomy (SOFIA). The receiver uses 2 $\times$ 7-pixel subarrays in orthogonal polarization, each in an hexagonal array around a central pixel. We present the first results for this new instrument after commissioning campaigns in May and December 2015 and after science observations performed in May 2016 . The receiver is designed to ultimately cover the full 1.8-2.5 THz frequency range but in its first implementation, the observing range was limited to observations of the [CII] line at 1.9 THz in 2015 and extended to 1.83-2.07 THz in 2016. The instrument sensitivities are state-of-the-art and the first scientific observations  performed shortly after the commissioning confirm that the time efficiency for large scale imaging is improved by more than an order of magnitude as compared to single pixel receivers. An example of large scale mapping around the Horsehead Nebula is presented here illustrating this improvement. The array has been added to SOFIA's instrument suite already for ongoing observing cycle 4. }

\authorrunning{Risacher et al.}
\titlerunning{the upGREAT 1.9 THz multi-pixel spectrometer}
\maketitle
%

\section{Introduction}
	
	Far-infrared astronomy has seen an enormous breakthrough with the observations of the Herschel satellite, which was active between 2009-2013  \citep{pilbratt2010}.  Herschel revolutionized our understanding of many astrophysical processes (star formation, planetary science, etc.) and so far resulted in more than 1600 refereed publications as of early 2016. 

   Since 2011, observations at infrared wavelengths 0.3-1600 {$\mu$m}  can be performed with the NASA/DLR airborne observatory SOFIA\footnote{SOFIA is a joint project of NASA and the German Aerospace Center (DLR).
The aircraft is based at the NASA Armstrong Flight Research Center (AFRC)
facility in Palmdale, California which also manages the program. NASA Ames
Research Center at Moffett Field, CA, USA, manages the SOFIA science and
mission operations in cooperation with the Universities Space Research Association
(USRA) headquartered in Columbia, MD, USA, and the German SOFIA
Institute (DSI) at the University of Stuttgart.}, carrying a 2.5m telescope onboard a Boeing 747-SP aircraft \citep{young2012}. 
It currently operates a suite of six instruments available to the interested communities covering wavelength ranges not observable from ground-based facilities:  EXES \citep{richter2010}, FIFI-LS \citep{klein2010}, FLITECAM \citep{smith2008}, HIPO \citep{dunham2008}, FORCAST \citep{adams2010} and GREAT \citep{heyminck2012}. The HAWC+ instrument \citep{vaill2007} is currently in its commissioning phase in 2016. Among those instruments, the only one able to perform high resolution spectroscopy (above $10^{7}$ resolving power) is the GREAT instrument\footnote{GREAT is developed by the MPI f\"ur Radioastronomie and the KOSMA/Universit\"at zu K\"oln, in cooperation with the MPI f\"ur Sonnensystemforschung and the DLR Institut f\"ur Planetenforschung.}. The other instruments employ continuum detectors and only offer low- to mid-resolution capabilities (except EXES which can go up to $10^{5}$ resolving power).  High resolution is needed in order to resolve the spectral lines profiles, which in turn allows studying the gas kinematics in great detail. 

Until now, only single pixel receivers had been achieved in a handful of observatories for frequencies above 1 THz.  The Max-Planck-Institut f\"ur Radioastronomie in Bonn, together with the I. Physikalisches Institut der Universit\"at zu K\"oln and the Institute of Planetary Research, (DLR, Berlin), are building a new generation of multi-pixel high resolution spectrometers for the SOFIA NASA/DLR project,  the upGREAT THz arrays. They consist of dual-color array receivers (for different frequency bands), which will ultimately operate in parallel. The Low Frequency Array (LFA) is designed to cover the 1.8-2.5 THz range using 2x7-pixel waveguide-based HEB mixer arrays in dual polarization configuration and double sideband mode. The High Frequency Array (HFA) will perform observations of the [OI] line at ~4.745 THz using a 7-pixel waveguide-based HEB mixer array, in double sideband mode. We present here the results from the upGREAT/LFA commissioning, which is the first multipixel array for high resolution spectroscopy successfully built and deployed on a telescope for this frequency range. This paper complements the technical description of the instrument \citep{risacher2016} with information about the actual on-sky performance and dedicated observing strategies relevant to interested users.  

\begin{figure}
  \centering
  \includegraphics[width=9cm,angle=0]{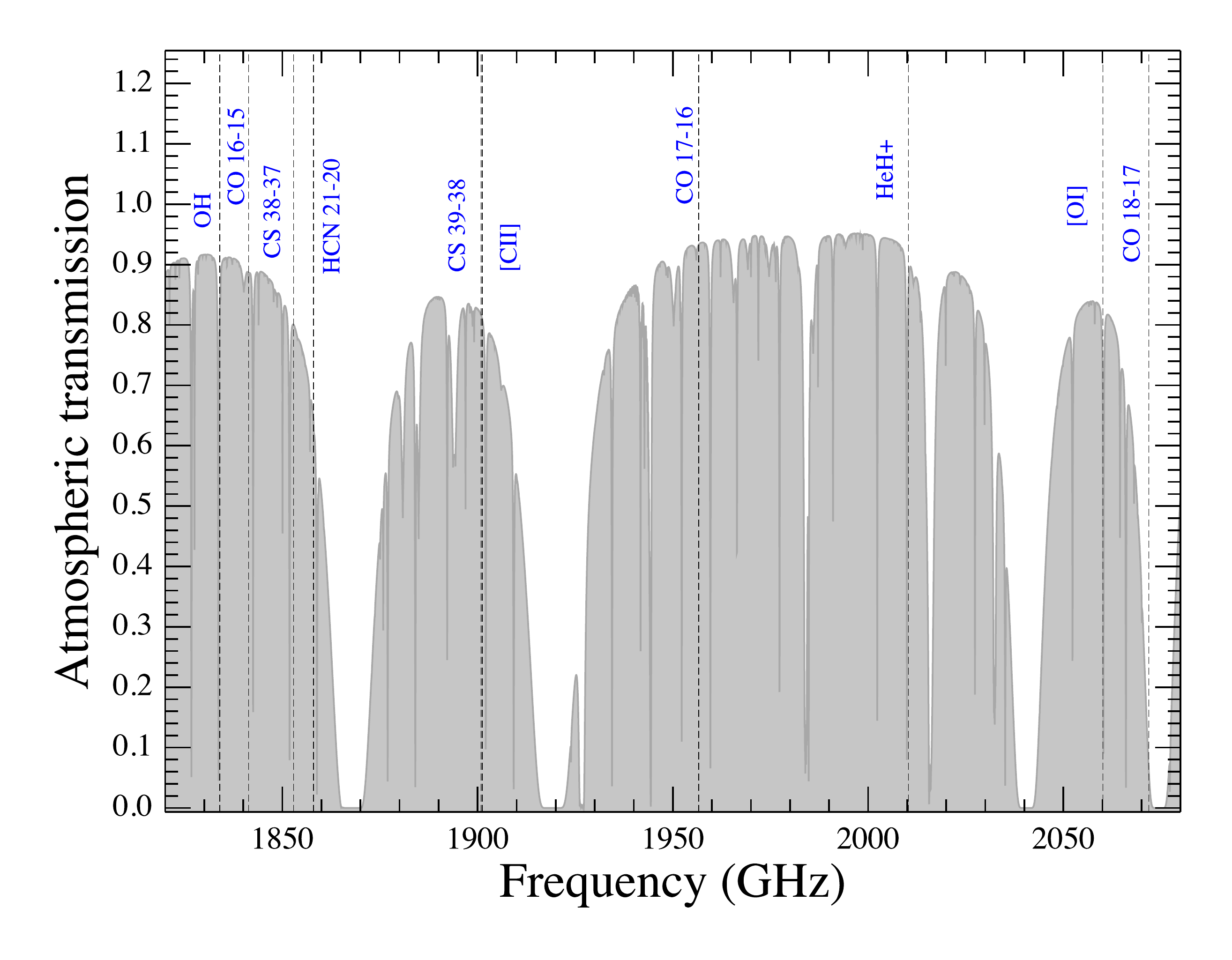}
  \caption{The upGREAT LFA receiver coverage for observing cycle 5 and atmospheric transmission for the SOFIA telescope when flying at an altitude of 40000 feet with a PWV of 15 {$\mu$m}. Important spectral lines are indicated.}
  \label{fig:atm}
\end{figure}

\section{Instrument Description}

\subsection{Frequency coverage and atmospheric transmission}

The far-infrared regime (1-10 THz) is largely blocked by the Earth's atmosphere.  Only very narrow windows with low transmission become accessible from dry high-altitude sites for frequencies above 1 THz.   To overcome the atmospheric attenuation, airborne telescopes, high-altitude balloons and satellites are the only platforms that allow performing such observations.   Figure \ref{fig:atm} shows the typical atmospheric transmission when SOFIA is flying at 40 000 feet, with average atmospheric conditions (residual Precipitable Water Vapour PWV of 15 $\mu$m).  The upGREAT/LFA receiver array is designed to cover the 1.8-2.5 THz frequency range, whose most important lines are listed in Table 1.  Among the main  atomic and molecular tracers are [CII], [OI], HeH$^{+}$, OH and CO via its rotational transition ladder from ${J=16-15}$ to ${J=22-21}$. For the commissioning and first flights, the tuning range was limited to the [CII] line transition at 1900 GHz by the available local oscillator (LO) reference signal. Since May 2016, the 1.83-2.07 THz range is  available. In the coming years it is foreseeable that with improved technology, the remaining of the tuning range will be covered.

\begin{table}[t]
  \centering
  \caption{Main atomic and molecular transitions accessible to upGREAT/LFA }
  \begin{tabular}{l l l l l}
    \toprule
    Species &  Rest Freq. & Transition & Comment \\
    \midrule
     OH & 1.834 THz &  & \\
     CO & 1.841 THz  &  ${J=16-15}$  &   \\
     CII & 1.9005 THz  &  ${J=3/2-1/2}$  &   \\
     CO & 1.955 THz  &  ${J=17-16}$  &   \\
     HeH$^+$ & 2.01 THz  &    &   \\
     OI & 2.06 THz  &   &   \\
     CO & 2. 07THz  &  ${J=18-17}$ &   close to atmospheric feature\\
     CO & 2.42 THz  &  ${J=20-21}$  &  not tunable yet \\
     OH & 2.49 THz & & not tunable yet  \\
     OH & 2.51 THz &   & not tunable yet\\     
     CO & 2.52 THz  &  ${J=21-20}$  &  not tunable yet \\
     \bottomrule
  \end{tabular}
  \label{tab:global}
\end{table}

\subsection{Instrument hardware}

	The upGREAT receiver is described in detail in  \citet{risacher2016}.   For high resolution spectroscopy at frequencies above 1 THz, the technology of choice for the detectors is based on Hot Electron Bolometers (HEB) superconducting mixers \citep{shurakov2016} and therefore, the instrument needs to be cooled to temperatures well below 6 K. Until now, such heterodyne receivers incorporated only up to a few mixers (e.g. \citet{graf2015}).  Therefore, integrating 14 such  mixers for the upGREAT/LFA  receiver represents a considerable technological challenge and a major step forward. All first generation GREAT receivers operate single mixers for frequencies between 1.2 THz and 4.7 THz \citep{heyminck2012}. The HEB mixers are developed and fabricated by the I. Physikalisches Institut der Universit\"at zu K\"oln  (\citet{putz2012}, \citet{buechel2015}). 

	 A view of the upGREAT/LFA cryostat is shown in Fig. \ref{fig:cryo2}, illustrating the dense packaging of the components. Fig. \ref{fig:great1} shows the whole instrument when installed on the SOFIA telescope. The LFA cryostat is cooled with a closed-cycle pulse tube refrigerator suitable for the superconducting detectors.   Table 2 lists the main receiver characteristics.   The receiver consists of two sub-arrays of 7 pixels each, having orthogonal polarizations, and placed in a hexagonal configuration around co-aligned central pixels.  
	 
The spectrometer backends are the last generation of Fast Fourier Transform Spectrometers (FFTS) \citep {klein2012} developed at the  Max-Planck-Institut f\"ur Radioastronomie, Bonn. They achieve an instantaneous  intermediate frequency (IF) bandwidth of 4 GHz per pixel, with a spectral resolution of 142 kHz, which translates to a total spectral bandwidth of about 630 \kms and a resolution of 0.022  \kms  at the frequency of the [CII] line.

\begin{table*}[t]
  \centering
  \caption{upGREAT LFA receiver characteristics}
  \begin{tabular}{l c l }
    \toprule
    Parameter &  Performance&  Comment \\
    \midrule
    RF Bandwidth  & 1.81-2.54 THz & Goal bandwidth  \\
                               & 1.83-2.07 THz & Currently usable RF bandwidth  \\    
    IF Bandwidth & 0.2-4GHz & full usable IF range (3dB roll-off)\\
    Receiver Sensitivity & 1800K SSB  & at 2 GHz IF (average over all pixels) \\
    System noise temperature& 2000K SSB & for PWV {$<$} 15 $\mu$m   \\
    Number of pixels & 7 per sub-array (14 total)  & Dual  polarization \\
    Backends & 4 GHz instantaneous BW & FFTS technology \\
    Array Geometry & hexagonal arrays with central pixel &  \\
    \bottomrule
  \end{tabular}
  \label{tab:global}
\end{table*}

\begin{figure}
  \centering
  \includegraphics[width=6cm,angle=0]{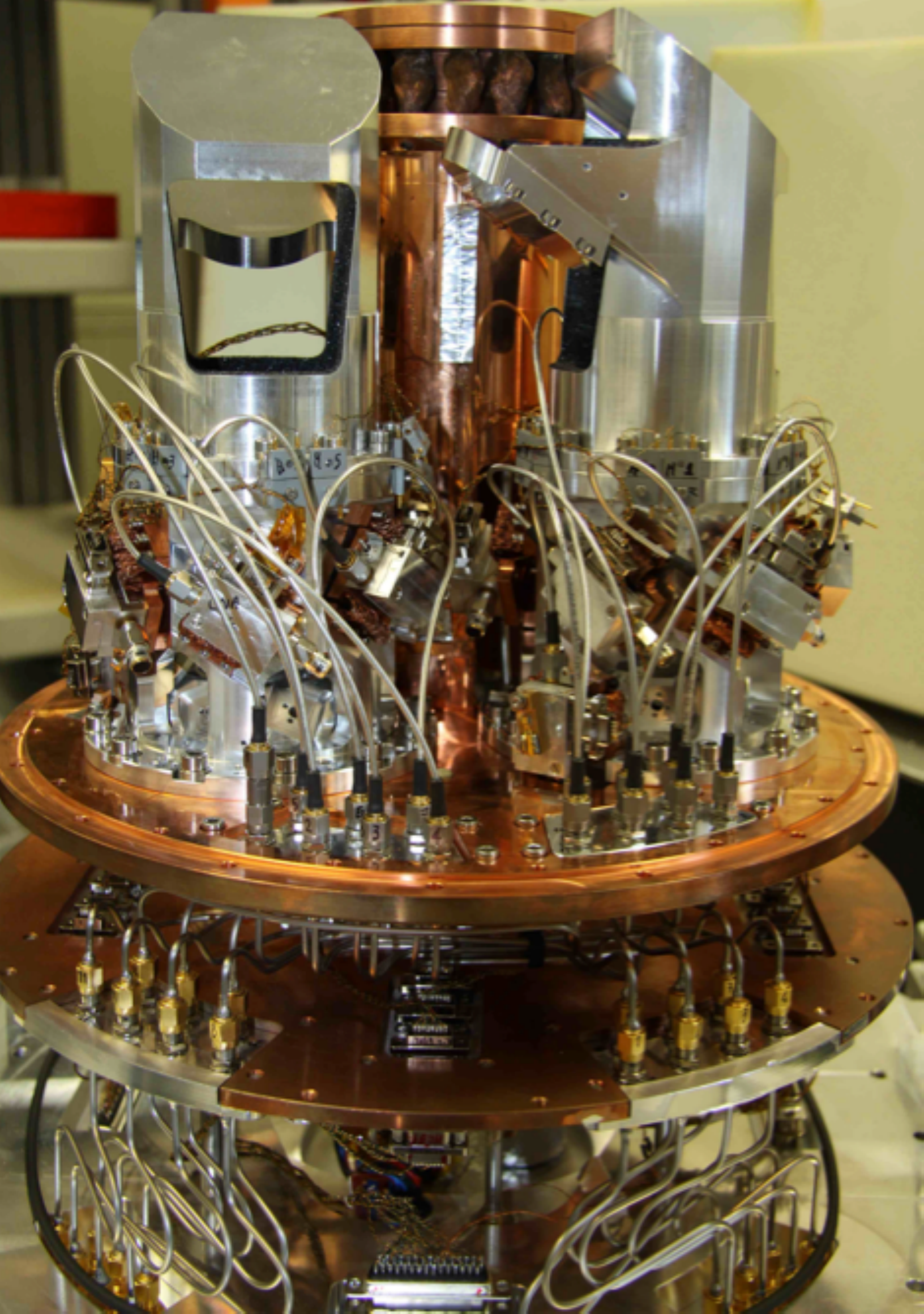}
  \caption{View of the inside of the upGREAT-LFA cryostat showing the different cold stages, part of the optics, and electronic connections.}
  \label{fig:cryo2}
\end{figure}

\begin{figure}
  \centering
  \includegraphics[width=8cm,angle=0]{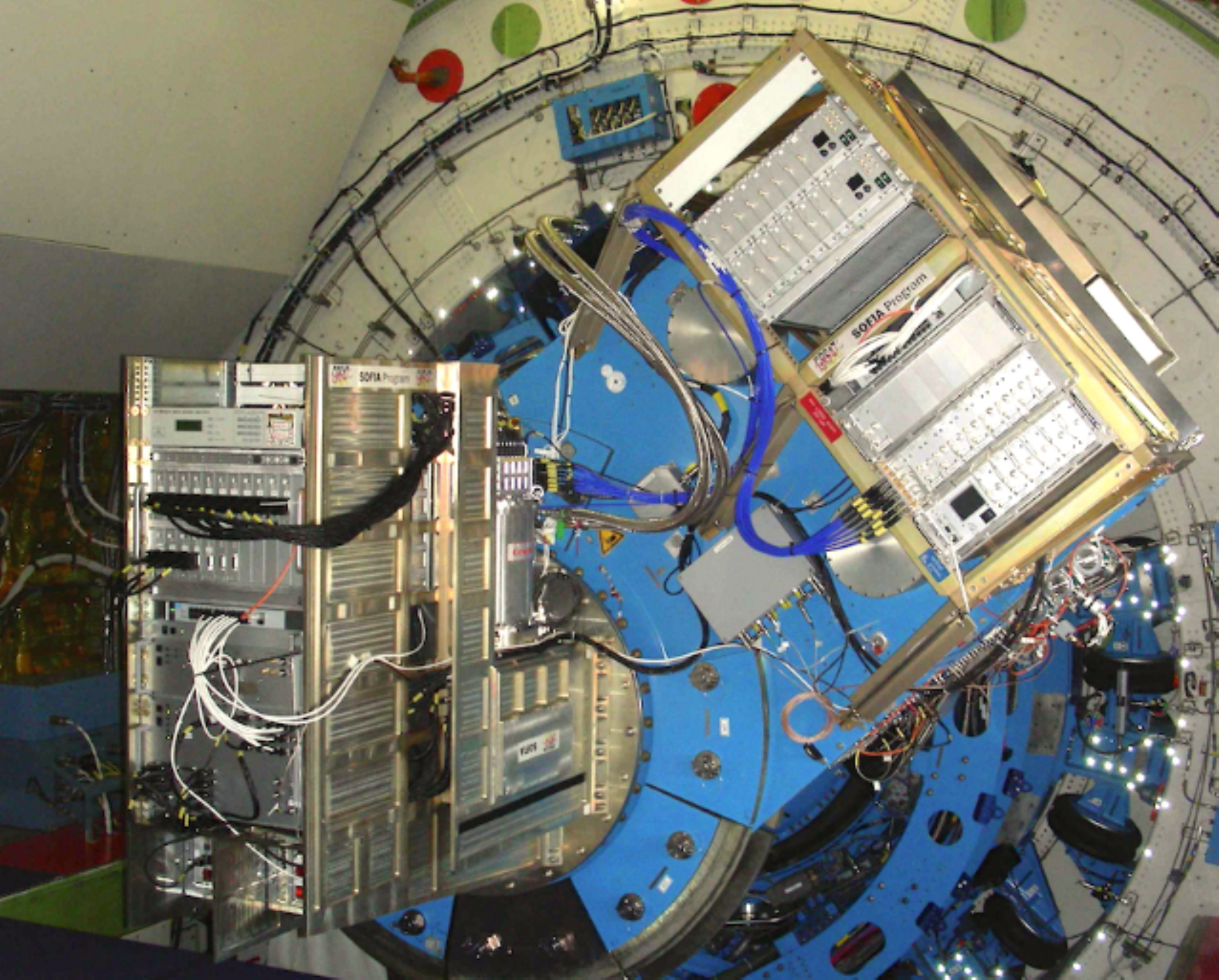}
  \caption{The upGREAT instrument mounted on the SOFIA telescope}
  \label{fig:great1}
\end{figure}

\subsection{Derotator}

With the upGREAT arrays, a derotator (implemented as a so-called K-mirror) is needed to compensate for the sky rotation in the instrument focal plane, which is fixed to the telescope (see Fig. \ref{fig:kmirror}). 
The SOFIA telescope is a 3-axis telescope, but has a limited travel range of only about $\pm 3^\circ$ in all axes for its fine drive. The telescope tracks celestial objects fixed in an inertial reference frame, keeping the relative orientation between the sky and the instrument focal plane fixed. When it gets close to the fine drive limit, the telescope is rewound to move close to the opposite limit, thus rapidly rotating the sky/instrument mounting flange focal plane orientation. The derotator then is turned oppositely, under computer control, by the same amount to keep the sky orientation in the instrument internal focal plane constant. 
The pixel pattern can then be kept fixed on sky (in the equatorial or horizontal reference system) and/or rotated as desired. Areas on the sky can therefore be efficiently homogeneously sampled, and a variety of observing strategies is possible.

\begin{figure}
  \centering
  \includegraphics[width=6cm,angle=0]{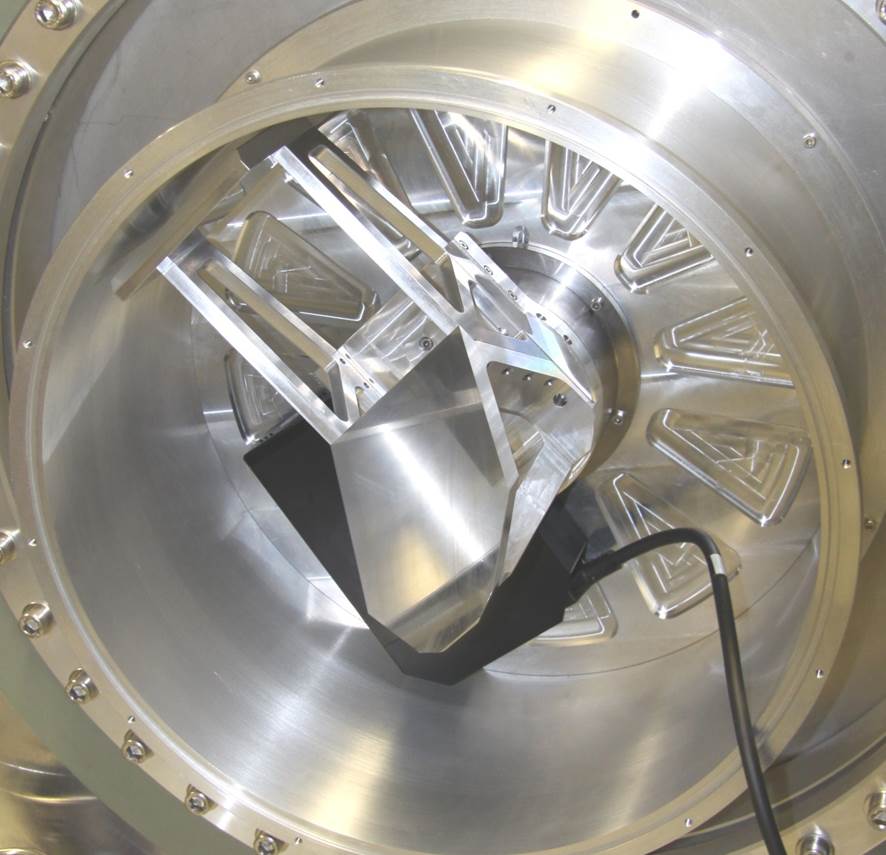}
  \caption{The derotator consisting of three high quality optical flat mirrors, mounted on a rotation stage, can compensate for the sky rotation to keep the array pattern in the instrument internal focal plane constant. }
  \label{fig:kmirror}
\end{figure}

\section{Commissioning Flights}

\subsection{Commissioning plan}

 The LFA was fully commissioned during a total of 5 flights, in May and December 2015. An example of one of the flight plans is shown in Fig. \ref{fig:flightplan}.  For the upGREAT commissioning flights, the SOFIA airplane took off and landed in Palmdale, CA. The duration of the flights is typically about 10 hours, of which ~8.5 hours are science observations. 

During the first flights in May, the instrument could only operate one of the two 7 pixel sub-arrays at a time. For the December flights, this limitation was resolved and the 2x7 pixels were employed simultaneously.  The activities performed during the commissioning were aimed at verifying: 

\begin{enumerate}[label=(\alph*)]

      \item basic functionalities of the new hardware/software (incl. the K-mirror derotator)
      \item the array performance on sky, i.e sensitivities and stability of the instrument
      \item determining the instrument focal plane geometry (boresight and pixels position on the sky)
      \item  assessing the telescope optimal focus position
      \item measuring the beam efficiencies on strong compact continuum objects
      \item testing and evaluating various observing modes
      
\end{enumerate}

\begin{figure}
  \centering
  \includegraphics[width=8cm,angle=0]{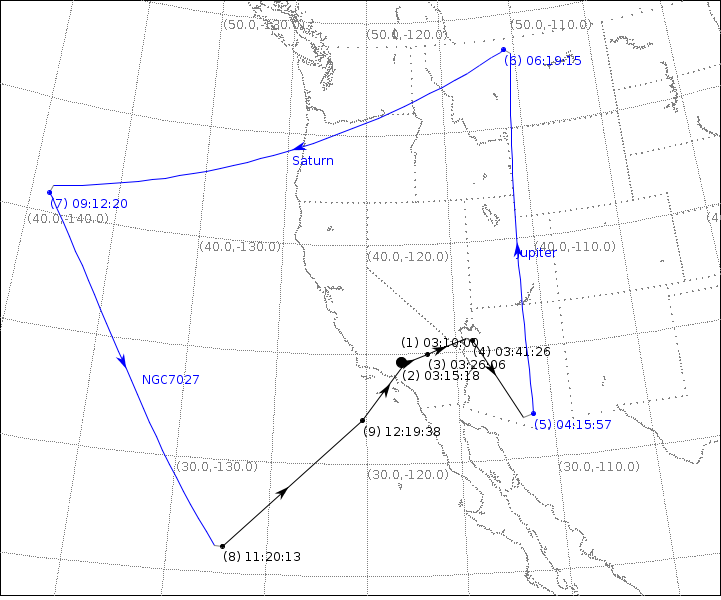}  
  \caption{Example of one of the LFA commissioning flights, performed on May 12$^{th}$ 2015. Indicated are the various targets observed during each of the flight legs, with Jupiter, Saturn and NGC7027 used for the commissioning.}
  \label{fig:flightplan}
\end{figure}

\subsection{Observing modes}

Observing with upGREAT on SOFIA is done with the \textit{kosma-control}-software package, 
developed and used for several single-dish telescopes and instruments; it is also used by the FIFI-LS instrument on SOFIA. 
All observing modes that have been used for single pixel observations \citep{heyminck2012} were adapted for the upGREAT observations. In addition, we introduced two new observing modes specific for a hexagonal array. All observing modes work both for total power and chopped observations, the latter are performed by moving the telescope subreflector.

The first new observing mode  is the array-optimized On-the-Fly (OTF) mapping mode (array-OTF mode). As shown in Fig.~\ref{fig:array_otf_scan}, scanning at 19.1 degrees against the axis of the hexagonal array makes an equally-spaced OTF scan with 7 pixels. To achieve a fully-sampled map, the array is scanned twice with a shift of 5.5\arcsec\ perpendicular to the scan direction. In order to reduce unwanted systematic effects due to different sensitivities between pixels, two approaches to achieve redundant coverage are recommended. The first one consists in performing two orthogonal scans. The other one consists in rotating the array by 60 degrees or integer multiples thereof. In this case, the central pixel scans the same line on the sky. In science observations, this array-OTF mode is optimal for large-area mapping. 

 A dedicated observing mode has been developed for the  the pointing measurements (array-pointing mode) in order to determine the position of each pixel in the instrument internal focal plane and the rotator axis position in the instrument mounting flange focal plane (see 3.4). To determine those parameters, at least two scans with different scanning directions should go through each pixel for one beam rotator angle, and this set of observations should be repeated for at least three different rotator angles. This array-pointing mode consists of a set of scans, each of which is defined by two pixels that should hit the source (see example in Fig.~ \ref{fig:array_pointing_scan_round1}.

\begin{figure}
  \centering
  \includegraphics[bb=70 110 420 260,width=\hsize,clip]{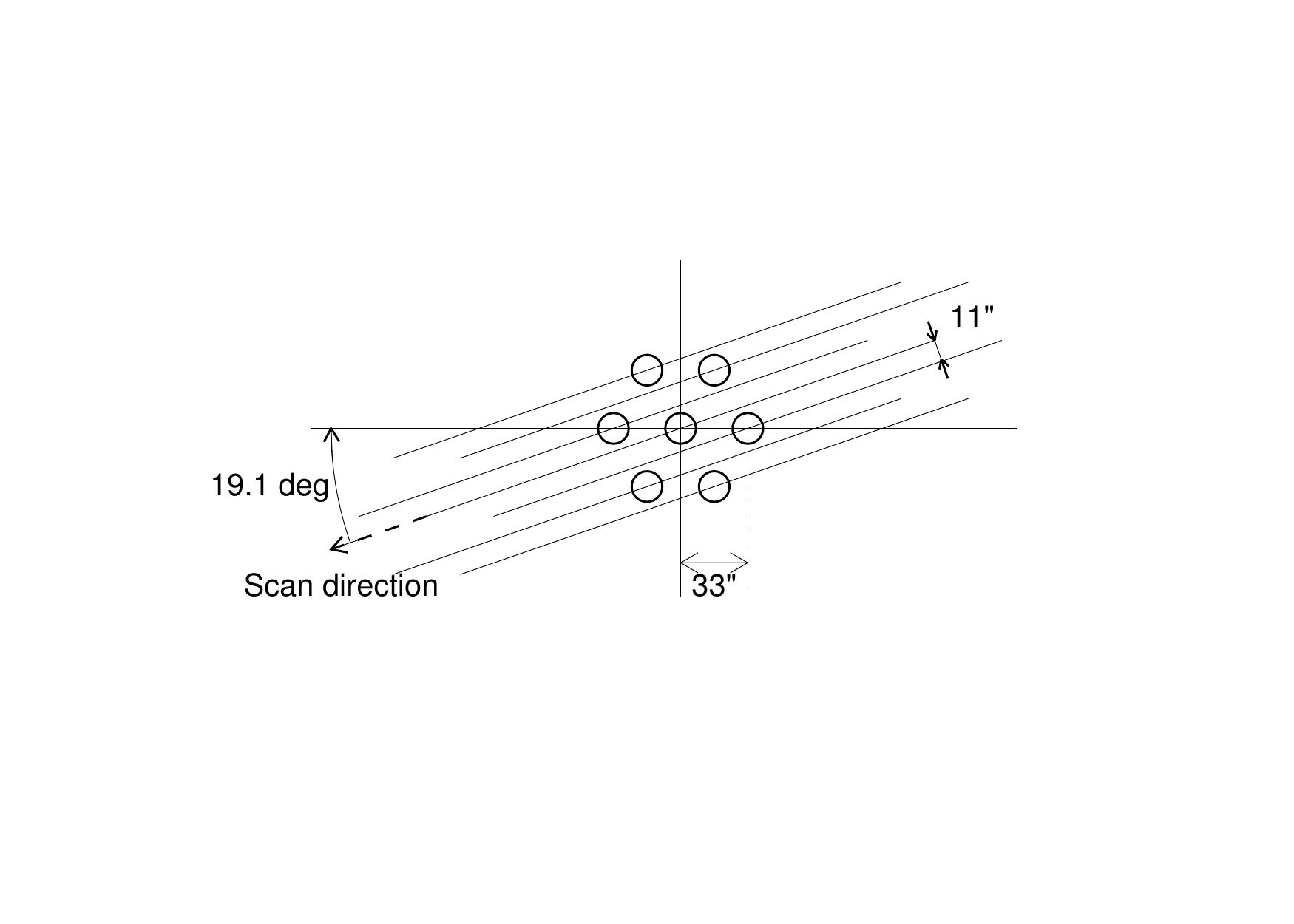}
  \caption{Sketch of the array-OTF mapping mode. Circles indicate the half power beam-width (HPBW) in the hexagonal array pattern. The pixel spacing is 33\arcsec.}
  \label{fig:array_otf_scan}
\end{figure}

\begin{figure}
  \centering
  \includegraphics[width=9cm,angle=0]{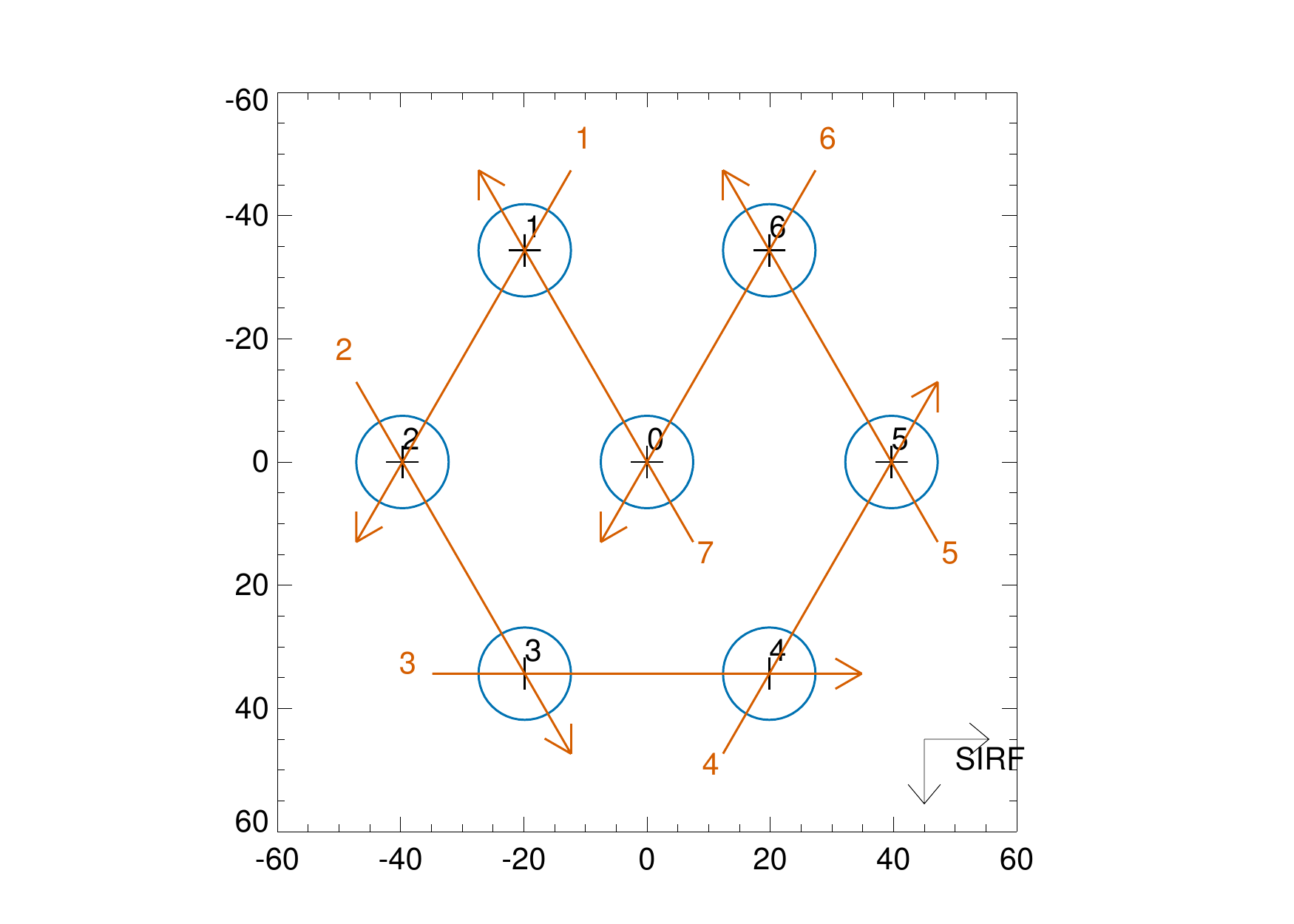}
  \caption{Example of the pattern of the scans in the array-pointing mode. The unit of the x- and y- axes is in arcsec, and the direction of the axis follows the science instrument reference frame (SIRF), which is the coordinate system that we use for the pointing measurements. The black crosses with blue circles indicate the pixels with their HPBW of 15\arcsec, and the red arrows indicate the scan direction.}
  \label{fig:array_pointing_scan_round1}
\end{figure}

\subsection{Instrument performance}

\subsubsection{Instrument sensitivities}

The receiver sensitivities were verified during the flights with the regular calibration sequences, typically performed every 5 minutes. The instrument looks sequentially at a cold load and a hot load absorber. The cold load absorber is at a temperature of about 70 K, and is located in a small closed-cycle cryostat which is cooled with a Stirling cooler. The transmission of its high density polyethylene (HDPE) window was carefully characterized across the RF frequency range (transmission at 1.9 THz is about 84\%). The hot load temperature is at about 293 K.   An example of the instrument receiver noise temperatures at the  [CII] frequency is shown in Fig. \ref{fig:trec}.   Typically, the receiver noise temperatures are about 600K-1000K double-sideband (DSB) at the lower IF range and increasing to 1200K-1800K DSB at the high end of the IF bandwidth (4 GHz). During the flights, the typical precipitable water vapour was in the range 10-15 $\mu$m, which then results in a system temperature  about 10-20\% higher than the receiver temperatures shown, typically below 2000 K single-sideband (SSB) at the IF of the line of interest. Depending on the available LO power, large parts of the 1.83-2.07 THz range have comparable receiver performance. For some tunings, specially close to the edges, only one of the polarizations receives sufficient LO power.

\subsubsection{Spectral purity}

With the current local oscillator chains, there are some very narrow spurs in the IF passband located typically between 0.8 and 1.3 GHz and between 3.0-3.8 GHz. The spurs are narrower than the FFTS channel width. Those single channels can be removed easily during post-processing of the data.  The positions of these interferences are not frequency dependent and they probably originate from electronic pickup noise in the multipliers bias lines.

\subsubsection{Baseline quality}

Baseline qualities of double beam chopped observations (the chop frequency is 2.5 Hz) are very good, and generally require removal of zero or first order polynomials only. In case of background continuum, the quality is limited by the precision of removal of residual atmospheric features.  The spectroscopic Allan variance time of the receiver channels is typically above 40 seconds, hence for long integrations per phase (i.e. position switched or long on-the-fly slews), some residual standing waves might be expected. The problem is exacerbated in observations asking for large offsets (at different airmass).  Compared to the Herschel/HIFI bands 6 and 7 which used HEB mixers and whose baselines suffered from strong standing waves , there is a tremendous improvement in baseline quality with the upGREAT/LFA.

\begin{figure}
  \centering
  \includegraphics[width=9cm,angle=0]{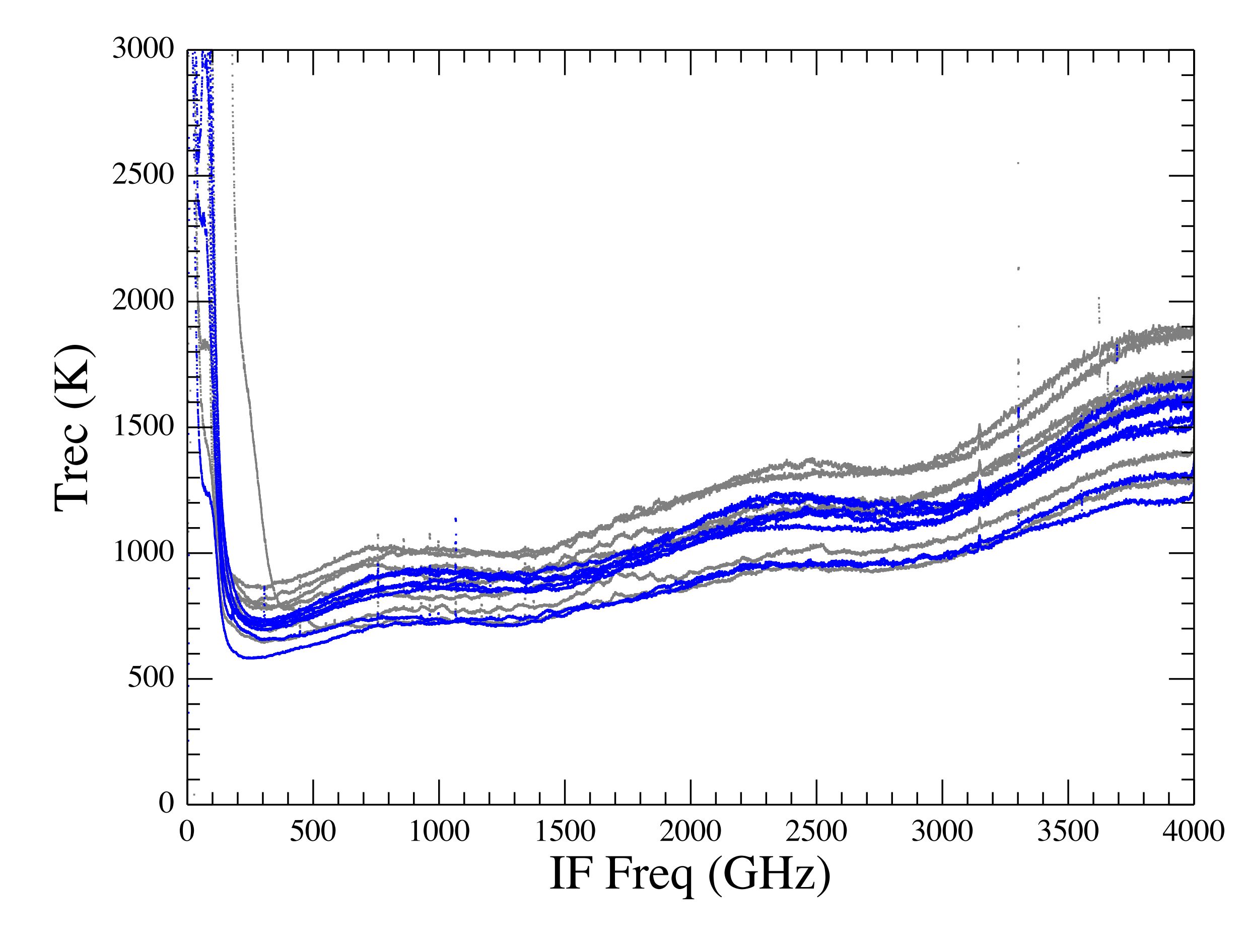}

  \caption{Receiver performance showing the noise temperature for all 14 pixels. The typical receiver noise temperatures are about 600K-1000K DSB at the lower IF range and increase to 1200K-1800K DSB at the higher end of the IF bandwidth (4 GHz). The seven dark blue curves are for one of the sub-arrays (H-polarization) and the seven light gray curves show the other one (V-polarization)}.
  \label{fig:trec}
\end{figure}

\subsection{Boresight determination}

The boresight determination is possible without a-priori knowledge of the exact array geometry. However, the relative positions of the pixels in the focal plane can be determined in the laboratory with a good accuracy \citep{graf2016}. This gives an estimate of the pixel positions on-sky, which will be corrected with pointing observations. The left plot of Fig. \ref{fig:focal} shows the derived pixel positions in the focal plane when measured in the laboratory prior to the installation on SOFIA in May 2016. 

The observations were done with the array pointing mode described in 3.3 (see Fig. \ref{fig:array_pointing_scan_round1}). For these measurements, a bright compact source like a planet, is needed. In May 2015, Jupiter and Saturn were used, and in December 2015, as a planet was not available at the beginning of the observations, a strong barely resolved spectral line source was used (the young planetary nebula NGC7027). For this, in addition to a cross scan pattern, a full map that includes the source in all pixels was used. This was the first time that a full map in the [CII] line was used to determine the array geometry. The measurements were verified with observations of Jupiter, available later in the flight series and subsequently used for the determination of the beam efficiencies.  In May 2016, Jupiter was used for the boresight determination.  

We parametrize the focal plane array geometry by the center offset from the derotator axis, and the scale-size and tilt of the ideal hexagonal footprint of each subarray. In a first step we thus fit this set of parameters to the set of measured pixel positions in the instrument mounting flange focal plane for at least three derotator angles. In a second step, we fit the offsets of each pixel relative to its nominal position in the ideal hexagonal subarray footprint to the set of measured positions. The former ones, i.e.\ the nominal pixel positions, are used to position the array and derotator angle when setting up on-the-fly mapping observations; the latter ones are used to record the proper, unavoidably slightly imperfect, pixel positions for the given derotator angle at the time of the observations in the raw data header information.

Both the  array-pointing mode and full map resulted in rms deviations of the individual pixel positions of 
$\sim$0.22~mm around the nominal positions in an ideal hexagonal pattern, corresponding to 0.89\arcsec on the sky (right plot of Fig. \ref{fig:focal}), confirming that the original laboratory measurements were very close to the true pixel positions.
The array central pixels were co-aligned within 0.5\arcsec.   The single pixel receiver L1 (1.25-1.5 THz) is used simultaneously with the upGREAT/LFA receiver and the relative alignment between the L1 and the central pixels of the upGREAT LFA arrays was verified to be within 4\arcsec. The central LFA pixels were used for the tracking reference, therefore 4" offset for the L1 beam is tolerable.

\begin{figure} 
  \centering
  \includegraphics[width=9cm,angle=0]{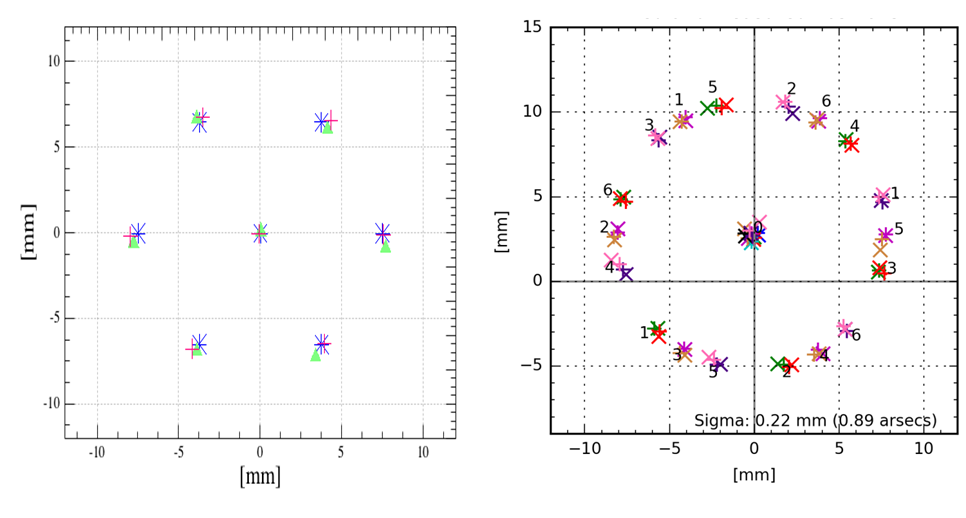}
  \caption{The left plot shows the derived focal plane pixel positions from laboratory measurements for both LFA sub-arrays (green triangles and red crosses), compared to the nominal designed positions (blue crosses). The right plot shows the on-sky derivation of the pixel positions, performed for different derotator angles. The subarray positions are plotted using a same color, for example red and green are the H and V polarizations subarrays at a given derotator angle, then the other colors are taken for different angles. The (0,0) for the plot refers to the SOFIA optical axis, which is different to the mechanical derotator axis. The rms deviations of the individual pixel positions of $\sim$0.22~mm around the nominal positions in an ideal hexagonal pattern, corresponds to 0.89\arcsec on the sky.}
  \label{fig:focal}
\end{figure}

\subsection{Beam Efficiencies}

The coupling efficiencies are determined from planets observations. Only  Jupiter and Saturn were available so far, in May and December 2015, and in May 2016. We present the May 2016 results, where all of the 14 pixels were successfully operational.  Jupiter was relatively extended (i.e.  $39.5\arcsec\times 37.0\arcsec$  on May 2016 relative to the half power full width of the diffraction limited beam of upGREAT: 15\arcsec at 1900 GHz) but well suited to determine and characterize the optical parameters of upGREAT.  The derived main beam efficiencies for the upGREAT array are typically 0.64 $\pm$ 0.03 and the spread between the various pixels can be seen in Table 3, and is of the order of 5\%.  Furthermore, for each pixel, these source was covered many times during the pointing measurements (9 scans), which provides an estimate of the measurement repeatability. The 1-$\sigma$ rms value is typically within 3\%. 

As the planetary models are typically accurate within $\pm{5}$\%, the derivation of the main beam efficiencies is estimated to be accurate within $\pm{8}$\%. The beam efficiencies are typically derived at the beginning of every flight series and the corresponding values are provided to the projects principal investigators (PI) together with the calibrated data.

\begin{table}[t]
  \centering
  \caption{upGREAT LFA beam efficiencies derived from Jupiter observations in May 2016 at 1900 GHz. The spread between pixels is of about $\pm{5}$\%. For each pixel, a 1-$\sigma$ rms level is indicated taken from 9 individual pointing measurements.}
  \begin{tabular}{c | c | c || c | c | c }
      \toprule
      Pixel &$\eta$$_{MB} $$_{(Jupiter)}$    & Pixel & $\eta$$_{MB} $$_{(Jupiter)}$  \\

    \midrule
    H0 & 0.67 $\pm$  0.03 &V0 & 0.64 $\pm$ 0.04 \\
    H1 & 0.63 $\pm$  0.02 &V1 & 0.64 $\pm$  0.03\\
    H2 & 0.57 $\pm$ 0.02& V2 & 0.61$\pm$ 0.03 \\
    H3 & 0.66 $\pm$  0.01&V3 & 0.65 $\pm$ 0.02\\
    H4 & 0.65$\pm$ 0.02& V4 & 0.59 $\pm$ 0.03\\
    H5 & 0.65 $\pm$ 0.03& V5 & 0.66 $\pm$ 0.03\\
    H6 & 0.63 $\pm$ 0.02& V6 & 0.63 $\pm$ 0.03\\

      \bottomrule
    Average & 0.64 & Average & 0.63 \\

  \end{tabular}
  \label{tab:global}
\end{table}

\begin{table}[t]
  \centering
  \caption{upGREAT LFA beam parameters at 1.9 THz}
  \begin{tabular}{l c c }
    \toprule
    Parameter &  Measured value \\
    \midrule
    Half Power Beam Width  & 15.1\arcsec ${\pm}$ 0.1\arcsec\\
    Pixel spacing  &  33\arcsec ${\pm}$ 0.2\arcsec\\
    Edge Taper (dB)  & 13 ${\pm}$ 0.2 \\
    Beam Efficiencies & 0.68 ${\pm}$ 0.03  \\
    Forward Efficiencies & 0.97 ${\pm}$ 0.01 \\
     \bottomrule
  \end{tabular}
  \label{tab:global}
\end{table}

\subsection{Calibration accuracy}

The calibration is based on the existing GREAT single pixel receiver calibration. The ${T_A^*}$ calibration scale of GREAT is based on frequent (typically every 5 min) measurements of our integrated hot- cold reference loads. Corrections for atmospheric absorption are done with the \textit{kalibrate}-task as part of the \textit{kosma-control} software package, which is described together with the detailed calibration scheme in \citet{guan2012}. 

In essence, a least square fit to all observed,  gain-calibrated sky-hot spectra (all positions and all frequencies) with the precipitable water vapour above the observatory as the fit parameter, is used together with an appropriate atmospheric model to determine the sky-transmission. The observed and fitted sky temperatures are delivered to the PI  with the data package, so they can appreciate the quality of the fit. 

The default implementation is to treat every upGREAT pixel as a separate signal and the derived opacities might converge to different values for each pixel.  The alternative scheme, which is being currently assessed, consists of using the pixels of each sub-array together and derive from them a best fit opacity value, which is then used for all pixels for the calibration. Observing with a receiver array, and hence obtaining independent measurements by each array pixel for the same atmosphere, allows to disentangle effects that are due to stability issues of the individual array pixels and effects from the atmosphere, that are common to all pixels. A detailed investigation of optimized calibration schemes is underway.  

By far the largest contribution to the calibration uncertainties stems from the atmospheric corrections. It is estimated to be accurate within $\pm{15}$\% when free of strong atmospheric features. The closer the observation is from a strong feature, the worse the accuracy.  The factors contributing to the overall calibration budget are: 

\begin{itemize}

      \item Atmospheric calibration: $\pm{15}$\%
      \item Hot and cold load blackbody temperatures: $\pm{2}$\%
      \item Sideband ratios: $\pm{5}$\%
      \item Standing waves, compression effects: $\pm{5}$\%
      \item Beam efficiencies: $\pm{8}$\%
      \item forward efficiencies: $\pm{2}$\%

\end{itemize}
 
Therefore the overall calibration error is estimated to be within $\pm{20}$\%.    The telescope pointing and tracking accuracy of 1\arcsec\   or better is close to perfect for our 15\arcsec\ beam.

\section{First science observations}

In May and December 2015, a total of five commissioning flights and five science flights were successfully performed.  A variety of projects could be observed during those flights. We present here a few snapshots that illustrate the large scale mapping capabilities of the array. These observations were targeting the [CII] line, which mainly traces the ionized and atomic gas and the UV illuminated edges (also called photo-dominated regions, PDRs) of dense molecular gas clouds. It is one of the main coolants of the warm gas and hence very important for the study of star-formation.

\subsection{W3OH }

The very first spectral footprint of 14 pixels of the LFA array was taken on the compact HII region W3(OH). This source is associated with one of the strongest bipolar outflows with a velocity range of about 50 km/s. The region hosts several sites of ongoing star formation. The most prominent ones are the ultra-compact HII region itself, which is ionized by a young O star, and a region showing hot molecular line and dust emission that is associated with strong water maser emission (W3(H2O)).  Fig \ref{fig:w3oh} shows the footprint of one of the subarrays in the [CII] line towards the source central position.  [CII] absorption by gas on the line-of-sight is recorded for the central pixel, towards the continuum of the hot core. The continuum level towards the hot core can accurately be determined in these double-beam chopped observations. As explained in 3.3.3, total power observations would not allow deriving a reliable continuum level. 

\begin{figure}
  \centering
  \includegraphics[width=9cm,angle=0]{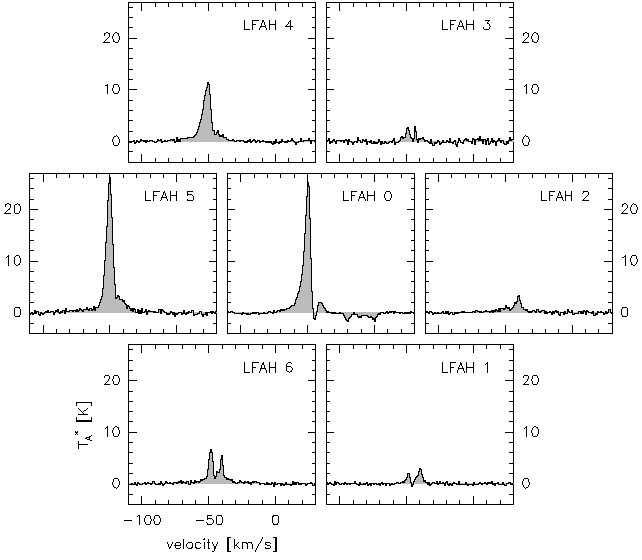}  
  \caption{Spectra for one of the upGREAT sub-arrays taken around the central position of the compact HII region W3(OH) (a polynomial baseline of order 0 was subtracted).  In the central pixel, towards the continuum of the hot core, [CII] absorption by gas on the line-of-sight is recorded.}  
  \label{fig:w3oh}
\end{figure}

\subsection{Horsehead large mapping demonstration }

After the successful technical commissioning of the LFA, the SOFIA Director granted discretionary time for a high-visibility science project to demonstrate the new opportunities for velocity-resolved spectroscopy to the SOFIA user communities. On SOFIA flight 267 on December 11, 2015, the [CII] emission from a large field towards the Horsehead nebula was recorded during an allocated time slot of four hours.  The field, 17.5\arcmin x12.5\arcmin in size, was fully sampled (6\arcsec step size) in total power on-the-fly observing mode. The field was substructured into four subfields, with redundant mapping (each subfield was covered four times with alternating scanning directions and rotated array geometry). Figure \ref{fig:horsehead} shows two maps showing the distribution of CII emission, integrated over the velocity ranges 9.5 to 11.5 \kms and 13 to 15 \kms. The sensitivity across the map is extremely uniform. The noise rms of the data cube, for a spectral resolution of 0.19  \kms, is ~2 K in the velocity channels free of emission.  

It is estimated that an equivalent map would have required over 50 hours of observing time using Herschel/HIFI, which would amount to a gain of more than a factor of 10 in mapping speed using SOFIA.The data are publicly available to the astronomical community as a demonstration observation with upGREAT.\footnote{\url{https://www.sofia.usra.edu/science/proposing-and-observing/proposal-calls/sofia-directors-discretionary-time/horsehead-nebula}}

The line/continuum intensities in the repeated subfields are typically reproducible within $\pm$5\%. When roughly comparing the integrated fluxes over similar regions observed with Herschel/HIFI (around the Horsehead) and taking into account the respective beam couplings, we found the fluxes to agree within 15\%,  which is within HIFI and upGREAT calibration error bars.

 \begin{figure}
  \centering
  \includegraphics[width=9cm,angle=0]{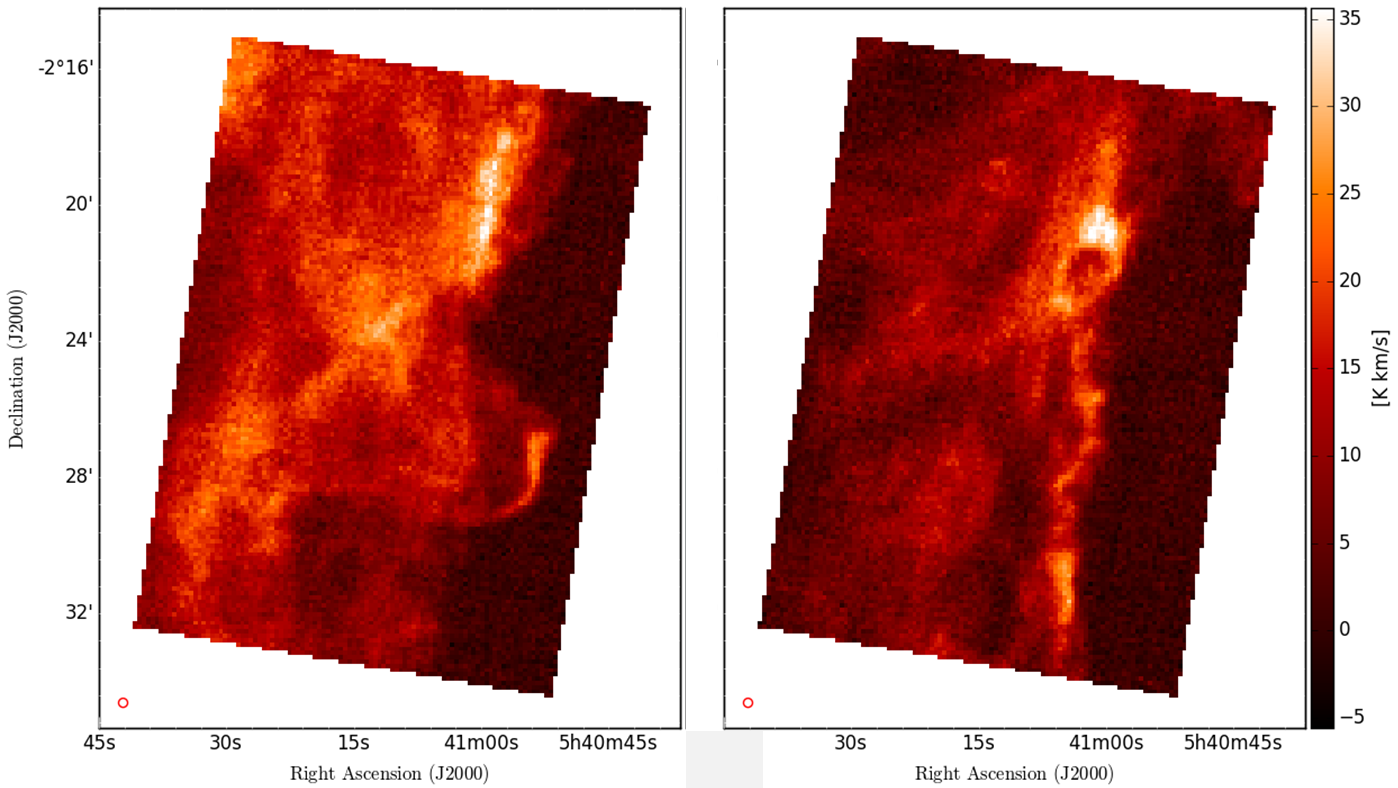}
 
  \caption{Example of large scale mapping of the Horsehead nebula region, performed in four hours, covering a $17.5\arcmin \times 12.5 \arcmin$ field. The lower left red circle show the 15" beam. We display two integrated channel maps at 9.5-11.3 \kms and 13-15 \kms, delineating the complex kinematics of the region. While strikingly prominent at optical wavelengths where it is seen as a dark patch, the Horsehead nebula [CII] emission is limited to the UV-illuminated edge of its PDR. The fine morphology of the IC434 PDR interface displays beautifully. }
  \label{fig:horsehead}
\end{figure}

\section{Future work}

We presented a novel multi-pixel spectrometer for observations at 1.9 THz, which was successfully commissioned onboard the SOFIA observatory in May and December 2015.  First results of large scale mapping of various sources demonstrate the capabilities of this instrument. With upGREAT on SOFIA, it is now possible to efficiently map very large fields with very high spectral resolution within a few hours, with excellent receiver sensitivities, homogeneous noise performance across the map and stable spectral baselines.
The next steps for this array will be to extend the usable RF bandwidth by utilizing a different local oscillator technology with sufficient power. 
A second multi-pixel receiver to observe the [OI] atomic fine structure transition at 4.745 THz, the upGREAT HFA, is currently under integration and will be commissioned at the end of 2016. The two upGREAT arrays will be operated simultaneously. 

\begin{acknowledgements}

The authors would like to thank the USRA and NASA staff of the Armstrong Flight Research Center in Palmdale and of the Ames Research Center in Mountain View as well as the DSI for their strong continuous support before and throughout the upGREAT LFA commissioning campaign. 
The development of upGREAT was financed by the participating institutes, by the Federal Ministry of Economics and Technology via the German Space Agency (DLR) under Grants 50 OK 1102, 50 OK 1103 and 50 OK 1104 and within the Collaborative Research Centre 956, sub-projects D2 and D3, funded by the Deutsche Forschungsgemeinschaft (DFG).
 \end{acknowledgements}

\bibliographystyle{aa} 
\bibliography{risacher} 

\Online
\appendix

\end{document}